\documentclass[useAMS,usenatbib, usegraphicx]{mn2e}
\usepackage{psfig}

\include{epsf}

\def\apj{{\rm ApJ}}
\def\apjs{{\rm ApJS}}

\def\etal{{\rm et~al.\ }}
\def\hmpc{\;h^{-1}{\rm Mpc}}

\def\hkpc{h^{-1}{\rm kpc}}
\def\kms{{\rm \;km\;s^{-1}}}

\def\kmsmpc{\kms\;{\rm Mpc}^{-1}}
\def\msun{{\rm M_{\odot}}}
\def\lya{Ly$\alpha$}
\def\lyb{Ly$\beta$}
\def\taueff{$\tau_{\rm eff}$}
\def\simlt{\lower.5ex\hbox{$\; \buildrel < \over \sim \;$}}
\def\simgt{\lower.5ex\hbox{$\; \buildrel > \over \sim \;$}}

\title[Lyman-alpha forest-CMB cross-correlation]
{
Lyman-alpha forest-CMB cross-correlation and the search for
the ionised baryons at high redshift
}

\author[R.A.C. Croft \etal]{
Rupert A.C. Croft$^{1,2}$\thanks{E-mail: rcroft@cmu.edu},
A. J. Banday$^{2}$ and Lars Hernquist$^{3}$\\
$^{1}$ Dept.   of  Physics,   Carnegie   Mellon  University,
Pittsburgh, PA 15213, USA\\
$^{2}$ Max Planck Institute for Astrophysics,
85741 Garching, Germany\\
$^{3}$ Harvard-Smithsonian Center for Astrophysics,
 60 Garden Street, Cambridge, MA 02138\\
}

\begin{document}

\date{\today}

\pagerange{\pageref{firstpage}--\pageref{lastpage}} \pubyear{2005}

\maketitle

\label{firstpage}

\begin{abstract}

The intergalactic neutral hydrogen which is responsible for the \lya\
forest of quasar absorption lines is a tracer of much larger amounts
of ionised hydrogen. The ionised component has yet to be detected
directly, but is expected to scatter CMB photons via the
Sunyaev-Zel'dovich effect. We use hydrodynamic simulations of a
LambdaCDM universe to create mock quasar spectra and CMB sky maps from
the same volume of space.  We find that the high redshift \lya\ forest
gas causes temperature fluctuations of the order of 1 ${\rm \mu}$K rms
in the CMB on arcmin scales.  The kinetic and thermal
Sunyaev-Zel'dovich effects have a similar magnitude at redshift three,
with the thermal effect becoming relatively weaker as expected at
higher redshift. The CMB signal associated with lines of sight having
HI column densities $ >10^{18} cm^{-2}$ is only marginally stronger
than that for lower column density sightlines. There is a much more
significant dependence of rms temperature fluctuation on mean \lya\
absorbed flux, however, suggesting that the CMB signal effectively
arises in lower density material.  We investigate the extent to which
it is possible to cross-correlate information from the \lya\ forest
and the microwave background to detect the Sunyaev-Zel'dovich effect
at redshifts $2-4$.  In so doing, we are enable to set direct limits
on the density of diffuse ionised intergalactic baryons.  We carry out
a preliminary comparison at a mean redshift $z=3$ of 3488 quasar
spectra from SDSS Data Release 3 and the WMAP first year data.
Assuming that the baryons are clustered as in a LambdaCDM cosmology,
and have the same mean temperature, the cross-correlation yields a
weak limit on the cosmic density of ionised baryons $\Omega_{\rm
b,I}$.  As a fraction of the critical density, we find $\Omega_{\rm
b,I} < 0.8$ at $95\%$ confidence. With data from upcoming CMB
telescopes, we anticipate that a direct detection of the high redshift
ionised IGM will soon be possible, providing an important consistency
check on cosmological models.

\end{abstract}

\begin{keywords}
Cosmology: observations -- large-scale structure of Universe
\end{keywords}

\section{Introduction}

The \lya\ forest seen in quasar spectra is now thought to be produced
by neutral hydrogen atoms in an intergalactic medium (IGM) which is in
photoionisation equilibrium (see e.g., Rauch 1998 for a review).  At
low to moderate redshifts, the fraction of neutral atoms is small (1
in $\sim 10^{6}$; e.g. Gunn \& Peterson 1965) but knowledge of the
photoionisation rate (from summing the contribution of known sources;
e.g. Haardt \& Madau 1996) makes it possible to extrapolate from this
one part in a million measurement to an estimate of the total baryon
density $\Omega_{b}$ (Weinberg \etal 1997, Rauch \etal 1997). Loeb
(1996) has argued that the ionised component of the IGM should be
directly observable by inverse Compton or Doppler scattering of cosmic
microwave background (CMB) photons by free electrons
(Sunyaev-Zel'dovich [1972,1980]; hereafter, the SZ effect).  Such a
detection would provide consistency checks on the currently favoured
cosmological model for structure formation, and place constraints on
the IGM density, temperature, ionisation state and velocity field.  In
this paper, we use a hydrodynamic cosmological simulation to
investigate how observations of the CMB temperature and \lya\ spectra
of quasars drawn from the same region of sky can be used in the search
for the hitherto unseen $99.9999\%$ of the hydrogen associated with
the \lya\ forest gas at redshifts $z \approx 2-5$.

In favoured models for structure formation, the optical
depth for \lya\ forest absorption arises in a continuously fluctuating
medium (e.g., Cen \etal 1994; Zhang, Anninos, \& Norman 1995;
Petitjean, M\"ucket, \& Kates 1995; Hernquist \etal 1996; Katz \etal
1996a; Wadsley \& Bond 1997; Theuns \etal 1998; Dav\'e et al. 1999).
The density of neutral hydrogen is inversely proportional to the
photoionisation rate $\Gamma$:
\begin{equation}
\Gamma=\int^{\infty}_{\nu_{HI}}d\nu\frac{4\pi J(\nu)}{h\nu}\sigma_{HI}(\nu) \, ,
\end{equation}
and the optical depth to absorption is given by
\begin{equation}
\tau_{Ly\alpha}=\frac{\pi e^{2}}{m_{e}c} f \lambda H^{-1}(z)n,
\end{equation}
where $f=0.416$ is the \lya\ oscillator strength, $n$ is the neutral
hydrogen number density, and $\lambda=1216$ \AA\ (Gunn and Peterson
1965).  In order for $\tau$ be of order unity, (as seen, for example at
redshifts $z\sim3$), $n$ must be much smaller than the total density
of hydrogen atoms, as noted earlier.  The gas in photoionisation
equilibrium lies at densities $\sim 1-10$ times the mean, and
obeys a power law relationship between density and temperature (Gnedin \&
Hui 1998). At $z=3$, according to hydrodynamic simulations (Dav\'{e}
\etal 2001), more than $90 \%$ of the baryons are expected to be in this
diffuse component. The tiny neutral fraction is readily detectable, but
not the dominant ionised portion of the material.  Owing to gravitational
evolution, at $z<2$ most of the baryons are expected to reside in a shock
heated IGM, with little neutral hydrogen (Cen \& Ostriker 1999, Dav\'e
\etal 2001).

The interaction of CMB photons with free electrons in the
intergalactic plasma was first considered by Sunyaev and Zel'dovich
(1972). Inverse Compton scattering preferentially increases the energy
of CMB photons, while conserving photon number, leading to a spectral
distortion whose amplitude is proportional to the product of electron
temperature and density (the thermal SZ effect). Doppler scattering
induces an intensity fluctuation with the same spectral shape as the
CMB itself (the kinetic SZ effect). The thermal effect is one of the
main sources of secondary CMB anisotropies on small angular
scales. 

The perturbation in the CMB thermodynamic temperature resulting from
scattering of nonrelativistic electrons is
\begin{equation}
\frac{\Delta T}{T}=y(x\frac{e^{x}+1}{e^{x}-1}-4)
\end{equation}
\begin{equation}
\simeq -2y \qquad {\rm for} \qquad x << 1,
\end{equation}
where $x=h\nu/kT_{\rm CMB}\simeq\nu/56.85$ GHz is the dimensionless
frequency and the second expression is valid in the Rayleigh-Jeans
limit, which we assume henceforth. The quantity $y$ is known as the
comptonization parameter and is given by
\begin{equation}
y\equiv \int dl \frac{n_{e}k(T_{e}-T_{\rm CMB})}{m_{e}c^{2}},
\end{equation}
where the integral is performed along the photon path.

The kinetic SZ effect arises from the motion of ionised gas with
respect to the rest frame of the CMB. The resulting temperature
fluctuation is $\Delta T/T=-b$ where
\begin{equation}
b\equiv\sigma_{T}\int dl n_{e} \frac{v_{r}}{c},
\end{equation}
gives the magnitude of the effect and $v_{r}$ is the component of the
gas peculiar velocity along the line of sight (positive sign for
receding gas, negative for approaching) to the observer.

The gas which causes \lya\ forest absorption has a density near the 
cosmic mean, with a volume weighted temperature around $2\times 10^4$ K at
$z=3$ (e.g, Schaye \etal 2000, McDonald \etal 2001). At this
temperature, we expect the contribution from the kSZ to dominate the
$rms$ temperature fluctuations of the CMB from the \lya\ forest.
Using observed \lya\ forest line counts, Loeb (1996) estimated that
the $rms$ temperature fluctuations from the kSZ would be $\Delta T/T
\sim 10^{-6}$ for the \lya\ forest lying between $z=2$ and $z=5$.
This is for angular scales of order 1 arcmin, with a $\theta^{-1}$
decrease on larger scales. There is much competition from other
signals on these small scales (including low redshift thermal SZ from
galaxy clusters, the Ostriker-Vishniac (1986) effect, patchy
reionisation (see, e.g. McQuinn et al. 2005), and foreground sources).
The suggestion to cross-correlate information from SDSS \lya\ forest
spectra and the CMB was also made by Loeb (1996) in order to try to
better extract the signal.  In this paper, we measure the kSZ effect in
simulations, as well as the thermal effect. The mass weighted
temperature of the IGM at $z\sim3$ is expected to be closer to $10^6K$
(e.g., Springel, White and Hernquist 2001), and so the thermal effect
should be significant. How much of that hot gas is physically associated
with \lya\ absorption will affect how well the tSZ can be detected by
cross-correlation.

We note that both being absorption phenomena, neither the SZ signal
nor the \lya\ forest opacity are directly affected by the inverse
square law. The problem of finding bright QSO background sources
aside, they therefore both have an advantage in the hunt for baryons
at high redshifts. Indeed, the value of the SZ effect for finding
galaxy clusters at the highest redshift has long been recognised (see
e.g., Carlstrom \etal 2002 for a review).  We aim to explore its
potential for finding diffuse ionised material at $z>2$.

The plan of this paper is as follows. In Section 2, we give details of
the hydrodynamic cosmological simulation and how mock \lya\ spectra
and CMB temperature sky maps were constructed from the same volume of
space. In Section 3, we compute statistical measures of the absorption
and examine correlations between the two probes of the intergalactic
medium. In Section 4, we compare directly to observational data from
the SDSS and WMAP, and in Section 5 we summarise and discuss our
results.

\section{Simulated spectra and skymaps}

\subsection{Hydrodynamic simulation}

We employ a hydrodynamical simulation of the popular $\Lambda$CDM
cosmology to make our prediction of the \lya\ and the SZ effects. In
order to both resolve the \lya\ features and the bulk velocities
responsible for the kSZ effect, we require a large simulation volume
with good mass resolution.  The data snapshots we use are from the
simulation ``G6'', also used by Nagamine \etal (2005) to which we
refer the reader for more details, which describes a representative
cubic volume of space $100 \hmpc$ on a side, in a cosmological
constant-dominated CDM universe, consistent with measurements from the
WMAP satellite (Bennett \etal 2003, Spergel \etal 2003).  The relevant
parameters are $\Omega_{\Lambda}=0.7$, $\Omega_{\rm m}=0.3$
$\Omega_{\rm b}=0.04$, and a Hubble constant $H_{0}=70 \kmsmpc$. The
initial linear power spectrum is cluster-normalised with a linearly
extrapolated amplitude of $\sigma_{8}=0.9$ at $z=0$.  The simulation
was performed with the smoothed particle hydrodynamics (SPH) code
Gadget-2 (Springel \etal 2001, Springel 2005), which is based on the
entropy-conserving approach of Springel \& Hernquist (2002), and is an
extension of the ``G-series'' runs of Springel \& Hernquist (2003b).

\begin{figure*}
\centerline{
\psfig{file=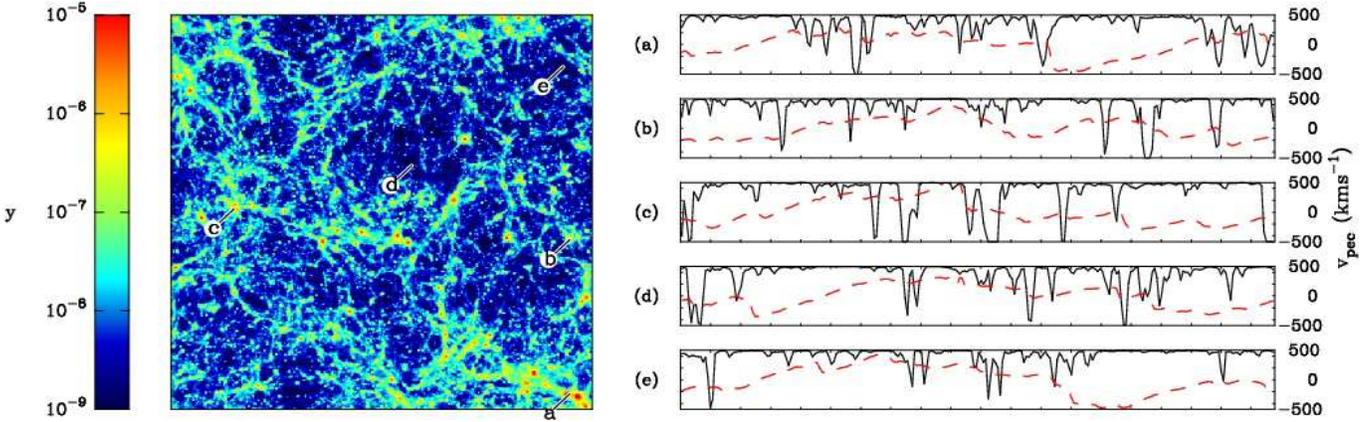,angle=-90.,width=18.0truecm}
}
\caption{
The tSZ CMB $y$ parameter (left panel) for a projection of one
simulation volume (side length 100 $\hmpc$). We show the positions of
five pixels for which we have drawn \lya\ sightlines (into the
page). The five spectra are shown in panels (a-e).  Black curves in
the right panels are the mock transmission spectra, while the dashed
red curves are the peculiar velocity.  Sightlines (a) and (b) and (c)
have been chosen so that $y$ is large (for (a), $y=5 \times 10^{-6}$,
for (b), $y=1.4\times 10^{-5}$), and for (c), $y=3.3\times 10^{-6}$),
all among the top 1 percent of $y$-values.  Sightlines (d) and (e) have
small values of $y$, $y=5.13 \times 10^{-10}$ and $3.8 \times
10^{-10}$ respectively.  Sightline (b) is the same as sightline (b) in
Figure \ref{mapspec} The \lya\ spectra are of length 9900 $\kms$ (one
simulation box length).
\label{mapspectsz}
}
\end{figure*}

The baryonic matter is initially represented by $486^{3}$ SPH
particles and the collisionless dark matter by $486^{3}$ N-body
particles. The initial mass per gas particle is therefore $9.7 \times
10^{7} \msun$ and dark matter $ 6.3 \times 10^{8} \msun$. The
gravitational softening length is $5 \hkpc$.  The calculation includes
gas dynamics, cooling, a multiphase treatment for star formation (see
Springel \& Hernquist 2003a), and a uniform background of ionising
radiation in the manner of Katz et al. (1996b), renormalising the
Haardt \& Madau (1996) spectrum to reproduce the observed \lya\ forest
mean optical depth.  A study of the SZ effect using this code was
carried out by Springel, White, \& Hernquist (2002), who noted that
the additional physics beyond adiabatic gas dynamics had a relatively
small effect; changing, for example, the mean comptonization by $20\%$.

\subsection{Mock \lya\ spectra}

We use simulation outputs from redshifts $z=2,3,4,5$ and $6$ to make
mock \lya\ spectra. For each output, this was done in the usual
manner, by integrating through the SPH kernels of the particles to
obtain the neutral hydrogen density field, and then convolving with
the line of sight velocity field (see e.g., Hernquist \etal 1996). At
each redshift, we compute $64^2$ lines of sight, equally spaced on a
square grid. In order to approximate longer spectra, we join together
spectra after performing a random translation, reflection and choice
of axis, as is done for example when making skymaps by ray-tracing
through multiple copies of the simulation volume (e.g., da Silva \etal
2000).  As our fiducial spectrum length, we take the distance between
the \lya\ and \lyb\ emission lines. Later in the paper, when we
cross-correlate the \lya\ forest and CMB at a particular redshift, our
analysis will refer to this \lya\ to \lyb\ region, centered on the
redshift in question. For reasons of computational simplicity, we
choose to ignore evolution over the length of each spectrum, both in
clustering and in the mean optical depth.  The \lya\ to \lyb\ regions
correspond to comoving spatial distances of $343, 343, 330$ and $314
\hmpc$ when centered on redshifts $z=2,3,4$ and $5$, respectively. We
add partial spectra to the full spectra to ensure the correct overall
length.

\begin{figure*}
\centerline{
\psfig{file=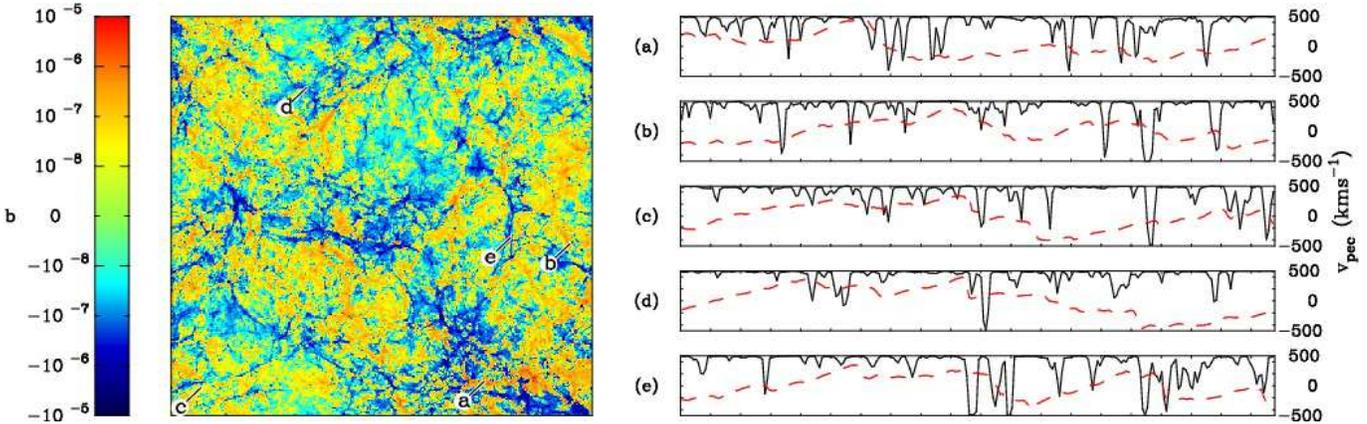,angle=-90.,width=18.0truecm}
}
\caption{
Projection through the simulation box showing the kSZ CMB 
Doppler parameter, $b$ (left
panel).  Pixels with a positive values of $b$ contain material moving
on average away from the observer, and negative $b$ values towards the
observer.  We show the positions of five pixels for which we have
drawn \lya\ sightlines (into the page). The five spectra are shown in
panels (a-e).  Black curves in the right panels are the mock
transmission spectra, while the dashed red curves are the peculiar
velocity.  Sightlines (a) and (b) have been chosen so that $b$ is
large and positive (for (a), $b=7 \times 10^{-6}$ and for (b),
$b=4\times 10^{-6}$), both among the top 1 percent of $b$ values.
Sightline (c) is the pixel with the smallest value of $|b|$,
$-8\times10^{-12}$. Sightlines (d) and (e) have large -ve values of
$b$, $-3 \times 10^{-6}$ and $-6 \times 10^{-6}$ respectively.  The
\lya\ spectra are of length 9900 $\kms$ (one simulation box length).
\label{mapspec}
}
\end{figure*}

For each spectrum, we compute the mean effective optical depth over
its entire length, i.e.,
\begin{equation}
\tau_{\rm eff}=-\ln \langle F \rangle,
\label{taueq}
\end{equation} 
where $\langle F \rangle$ is the mean absorbed flux in the spectrum,
$\langle e^{-\tau} \rangle$. Throughout most of this paper, we will use
this quantity, $\tau_{\rm eff}$, to determine the relative amount of
\lya\ absorption in each sightline. Another possible measure is to
compute the total column density of neutral hydrogen along each
sightline from \lya\ to \lyb\ . Being unaffected by saturation, this
will tend to reflect the presence of individual high density clouds.

As noted earlier, the simulation was run using the UV ionising
background of Haardt \& Madau (1996). In our subsequent analysis, we
rescale the \lya\ optical depths by a small amount after computing the
spectra by adjusting them slightly so that $\langle F \rangle$
averaged over all spectra reproduce the observational results of
Schaye \etal (2004), $\langle F \rangle=0.878,0.696,0.447$ for
$z=2,3,4$.

\subsection{CMB skymaps}

At each redshift, we construct maps of the Compton $y$ parameter and
the Doppler $b$ parameter by projecting the appropriate SPH kernel
weighted quantities along one axis. Unlike Springel \etal (2001), for
example, we do not project along lines which meet at a single $z=0$
observer, as we will be interested only in the SZ signal generated by
gas which falls within the same redshift interval as our \lya\ forest
spectra. In order to compare CMB maps and spectra for the same volume
of space, we randomly translate and reflect our SZ maps in the same
manner so they match up with the spectra from the previous section. We
generate 40 maps, which have an angular size which subtends the whole
box at each particular redshift; i.e., 94.7 arcmin at redshift
$z=2$, and 77.2, 68.5, 63.1 and 59.5 arcmin at $z=3,4,5$ and $6$
respectively. We make maps at a resolution corresponding to $2048^2$
pixels, but also average the pixels so that we have $64^2$, the same
number as the \lya\ spectra.

\subsection{Examples}

Looking at kSZ maps and \lya\ spectra generated from the same region
of space, we can see if there are any obvious visual relationships
between the CMB temperature in pixels and the appearance of the
absorption features. In Figure \ref{mapspectsz}, we show the Compton
$y$ parameter, from the thermal SZ effect at $z=2$, and alongside it
five spectra which are taken with the sightline along the axis
perpendicular to the plane of the map.  Here, in both maps and
spectra, we do not plot the entire \lya\ to \lyb\ region (we do this
later, below), but just the material in one single cubic simulation
volume.  It is apparent from the map that much of the volume
contributes to a very small $y$ parameter, and that most of this
redshift slice has $y < 10^{-9}$ or below. There are also a number of
obvious galaxy clusters, with $y >10 ^{-5}$

Of course, a map of the $y$ parameter from all material between $z=0$
and the last scattering surface would have quite a different
appearance, with the mean $y$ parameter being higher, of the order of
$10^{-6}$, and with many of the obvious filaments seen in Figure
\ref{mapspectsz} washed out by projection effects. Figure 2 of White
\etal (2002) demonstrates this, using a similar simulation.

Sightlines (a)-(c) in Figure \ref{mapspectsz} were chosen from those
among the $64^{2}$ we made which have large values of $y$. Spectrum
(a) is visibly taken from near a cluster, seen in the SZ map. There,
however, does not seem to be much evidence of this in \lya\
absorption, at least in terms of saturated absorption. The other two
spectra, (b) and (c) are also fairly similar, even though they
are both associated with large $y$ values.  Because there is not much
saturated absorption, we can see that it is unlikely that column
density will correlate well with SZ decrement.

Sightlines (d) and (e) in Figure \ref{mapspectsz} have been chosen to
have values of $y$ near the bottom of the rank-ordered list of all
pixels. From the SZ plot it can be seen that the spectra lie on
sightlines though obvious voids, and the $y$ values are 10000 times
smaller than for panels (a)-(c). On examination, these spectra appear
to have very few wide absorption features compared to spectra (a)-(c),
and the overall level of \lya\ absorption is slightly lower
(particularly (e)).

Turning now to the map of the kSZ effect (Figure \ref{mapspec}), we
see that much of this redshift $z=2$ slice has a Doppler $|b|$ value
of $\sim 10^{-8}$, although there are several clusters and filaments
dense enough and moving fast enough to reach $|b|=10^{-5}$. Clusters
can be seen moving towards the observer (blue) and away (red) . The
large coherence length of the velocity field can be seen also, from
the size of the red and blue structures.  Projecting the kSZ effect
for a larger volume, for example subtending $z=2-5$, will lead to more
structures but also more cancellation of positive and negative $b$
values. The general level of fluctuations likely agrees at least
roughly with the estimate of Loeb (1996) ($rms$ $b \sim 10^{-6}$).

Again, the kSZ plot can be compared to full lightcone maps (e.g. Figure
2 of Springel \etal 2001), which show a similar pattern of approaching
and receding material, also with less obvious filamentary
structure. Sightlines (a) and (b) in Figure \ref{mapspec} were chosen
to have large positive Doppler $b$ values, and (d) and (e)
negative. It can be seen that their spectra contain fairly substantial
amounts of absorption, and one can understand how the mass weighted
velocity will give rise to the $b$ values seen. Panel (c) has a
velocity field which cancels in the left and right halves of the
spectrum, yielding a $b$ parameter much closer to zero ($b \sim
-10^{-11}$).

The number of sightlines passing close to galaxy groups and clusters
in Figures \ref{mapspectsz} and \ref{mapspec} is relatively small, at
least in this redshift interval. This means that any clustering
statistic which weights spectra equally will tend to be sensitive to
physical conditions in the regions with relatively small overdensity.

\begin{figure*}
\centerline{
\psfig{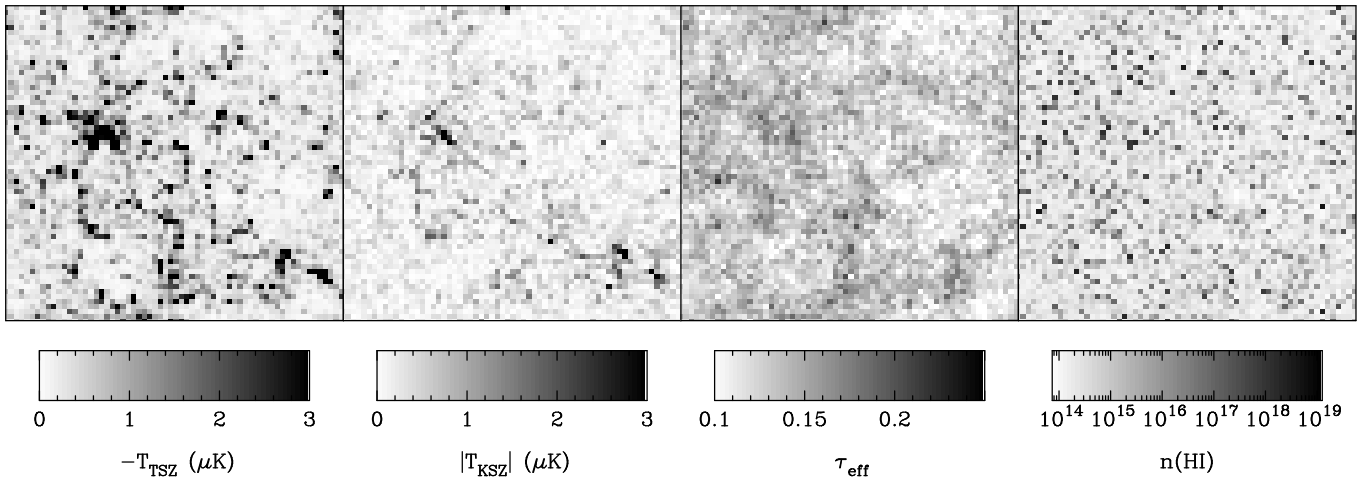}
}
\caption{
Sky maps of CMB temperature (left panel) owing to the kSZ effect at
z=2 (for a volume equivalent to the entire \lya\ to \lyb\ spectral width).
The width of the
panel is 95 arcmin.  We also show $|T|$, mean flux $<F>$ in \lya\
spectra and total column density in \lya\ spectra. In each map, there
are $64\times64$ pixels.
\label{maps}
}
\end{figure*}

Looking at a few individual sightlines, it seems likely that there is
a connection between \lya\ forest absorption and SZ-induced
temperature fluctuations.  One way to examine this further would be to
look at skymaps of both the \lya\ forest and CMB. Of course, with real
observations it unlikely that we would have a dense enough grid of
background quasars to make a true \lya\ map, but with the simulations,
we can use such a technique to examine the spatial pattern of
absorption.  In Figure \ref{maps}, we show examples of such maps,
plotting the thermal and kinetic SZ effects alongside angular maps of
the \lya\ optical depth and column density. It should be noted that we
plot the absolute value of the temperature change produced by the
kinetic effect.  Unlike the previous two figures, here we show a
projection through the volume of space corresponding to an entire
\lya\ to \lyb\ region (not just one simulation box). The results are
centered on redshift $z=2$, and the angular size of the maps is 95
arcmins. Because we have a grid of $64^{2}$ \lya\ spectra, we have
rebinned the CMB maps so that they have the same resolution (1.5
arcmin pixels).

The grayscales of the SZ maps are the same, and so we can see from the
left two panels of Figure \ref{maps} that the thermal and kinetic
effects trace similar features, with the thermal effect being more
prominent at this redshift. Because of the differing frequency
dependence of the two SZ effects, cross-correlation of the two maps
may eventually be possible using observational data, as suggested for
example by da Silva \etal (2000). In the present context, we are
interested in the potential cross-correlation of the \lya\ forest
spectra and the SZ effects. In the third panel of Figure \ref{maps},
we show the mean effective optical depth, given by Equation
(\ref{taueq}) for spectra through the center of each pixel. It can
clearly be seen that the patches of absorption largely follow the SZ
temperature well, although there are several regions with large
differences. The \lya\ \taueff\ fluctuations have a somewhat lower
contrast, varying by about a factor of two across the map. The obvious
groups and clusters in the SZ maps do not show up as prominently in the
\lya\ map. This is likely to at least partly owe to the fact that
regions with large amounts of \lya\ absorption are saturated.

This is not the whole story, however, as we can see by looking at the
rightmost panel of Figure \ref{maps}. Here, we show the HI column
density in each spectrum integrated from \lya\ to \lyb\ . There is
much smaller scale structure present, as the dense knots of neutral
hydrogen are picked out. The range of column densities varies from
$N_{HI}=10^{14}$ cm$^{-2}$ to $10^{19}$ cm$^{-2}$.  The structures in
$N_{HI}$ do not correlate well with the CMB maps. The smooth,
large-scale structures in the SZ effect maps (and \lya\ \taueff\ ) are
not readily seen. This observation will guide our use of \taueff\ as
the main \lya\ statistic to be used.

\begin{figure}
\centerline{
\psfig{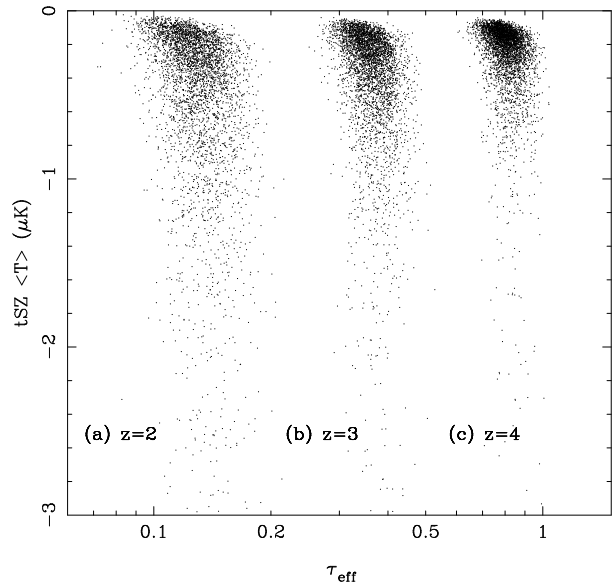}
}
\caption{
Thermal SZ temperature decrement $T_{tSZ}$ vs \lya\ forest $\tau_{\rm
eff}$ for spectra at three different redshifts.
\label{tvstau}
}
\end{figure}

In Figure \ref{tvstau}, we show a scatter plot of the \lya\ \taueff\
against the thermal SZ temperature decrement, at redshifts $2,3$ and
$4$, again from maps with $64^{2}$ pixels (1.5, 1.2 and 1.1 arcmin
size for $z$=2,3,4). The correlation between the two quantities is
apparent, and in particular at each redshift there are few lines of
sight with very low values of \taueff\ but substantial SZ
decrements. There is a large amount of scatter, however, indicating
that even if foreground/background noise is not an issue, it will be
necessary to average over many sightlines to obtain statistically
significant results.

\section{Statistical measures}

In this section, we use the set of skymaps and spectra generated from
our simulation to examine statistically the CMB fluctuations caused by
\lya\ forest gas.

\subsection{Rms temperature fluctuations}

\begin{figure}
\centerline{
\psfig{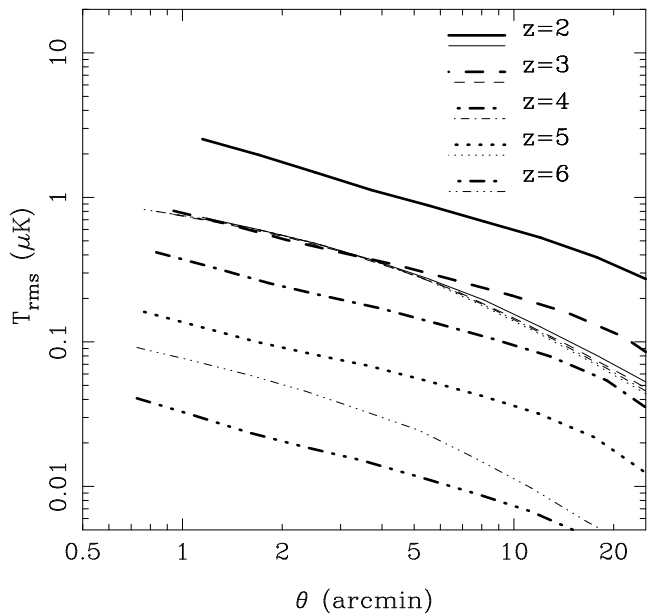}
}
\caption{
Rms CMB temperature fluctuations in top-hat circles as a function of
radius (x-axis) for three different redshifts. In each case, we plot
the rms fluctuations associated with a volume which would contain
material with HI absorption wavelengths from \lya\ to \lyb\ centered
on the indicated redshift. Thin lines are kinetic SZ and thick lines
thermal SZ.
\label{rmsksz}
}
\end{figure}

In Figure \ref{rmsksz}, we show the $rms$ temperature fluctuations
averaged in top-hat circles on the plane of the sky, calculated from
maps such as those shown in the left two panels of Figure
\ref{maps}. We show results centered on integer redshifts between
$z=2$ and $z=6$, in each case, calculating only the CMB fluctuations
caused by gas within the \lya\ to \lyb\ region of the spectrum. The
thermal and kinetic SZ effects are shown separately, and it can be
seen that at $z=2$, $T_{rms}$ is approximately three times larger for
the tSZ effect, as we expect from Figure \ref{maps}. At $\theta=1 $
arcmin, the $rms$ $T$ fluctuations are $\sim 2 \mu$K for the tSZ at
this redshift.

The curves for all redshifts have an angular dependence given by $\sim
\theta^{0.6}$.  At $z=3$, the two SZ effects are comparable, and
beyond that the kSZ effect stays approximately constant out to $z=5$.
The $rms$ peculiar velocity of the gas increases slightly with
decreasing redshift, as detailed in Table \ref{tabz}. This is balanced
by the decrease in angular size for a given comoving length which
shifts the curves to the left.  The main change in the kSZ at $z=6$ is
caused by the switching on of the UV ionising background in our model,
and so the ionised fraction is small then.  The temperature of the gas
influences the fluctuations owing to the thermal effect directly, and
this has a much more rapid dependence on redshift. In Table
\ref{tabz}, we can see that at $z=6$, the electron-weighted mean
temperature is $>10^{5}K$, as the free electrons are concentrated in
the few collisionally ionised structures present at that
redshift.\footnote{We ignore any difference between the electron and
ion temperatures; see Yoshida et al. (2005).}  By $z=5$ when most of
the IGM is photoionised, the electron-weighted temperature is $2.6
\times 10^{4}$ K, and as gravitational evolution continues, shock
heating increases the temperature by another factor of 20 by
$z=2$. The rise in $T_{rms}$ owes mostly to this, with a small part
coming from the factor 2.3 increase in linear density fluctuations
themselves between the two redshifts.

We note that although the thermal SZ effect is dominant at redshifts $z<3$,
the signal that is actually generated in gas that 
directly produces the \lya\ forest is a small fraction of the
total thermal  SZ signal at that redshift. This is because the
temperature of \lya\ forest-producing
gas at the mean density at $z=3$ is $\sim 10^{4}$K, low
enough that the thermal SZ effect should be extremely small (e.g., Loeb 
1996). The thermal SZ 
signal we are instead seeing is largely generated in
the Warm-Hot Intergalactic Medium (WHIM, e.g. Dav\'{e} \etal 2001)
at these redshifts which
is collisionally ionised, and is responsible for the higher mass-weighted 
temperature seen in Table 1. As we discuss below, the thermal 
SZ signal at this redshift is strongly correlated with the \lya\ absorption
signal, indicating that the WHIM signatures are physically 
associated with the same large-scale structures as the \lya\ forest,
tracing their higher density regions.

\begin{table}
\caption{The mass density and electron number weighted gas temperature and 
velocity dispersion in the simulation at different redshifts.}
\begin{center}
\begin{tabular}{ccccc}
$z$  & $T_{\rho}$ &  $T_{e}$  & $v_{rms,\rho}$ & $v_{rms,e}$\\  
  &(K) &	(K)	&$(\kms)$&	$(\kms)$\\
		\hline
6&    $7.1\times 10^3$&   $1.9\times10^5$	&136 &	 166\\ 
5 &   $2.57\times 10^4$&	$2.60\times 10^4$  &149&	 148	\\		
4&	$6.86\times 10^4$&	$6.89\times 10^4$	&164&	164\\
3&	$1.77\times 10^5$&	$1.76\times 10^5$	&185&	185	\\
2&	$5.04\times 10^5$&	$5.04\times 10^5$	&214&	214\\
\hline
\end{tabular}
\end{center}
\label{tabz}
\end{table}

\subsection{Cross-correlation function}

We have seen that IGM gas at redshifts $z=2$ to $z=5$ yields rms CMB
temperature fluctuations on the order of a few $\mu$K. One could
calculate the angular power spectrum, the $C_{l}$s, and then see
whether it could be separated from other small angle
fluctuations. Owing to the multitude of other possible signals,
however, it is likely to be more useful to cross-correlate with
material at a known redshift, and use this to extract information. As
the \lya\ forest contains most of the baryons at these redshifts, and
the mean effective \lya\ optical depth appears to correlate visually
with CMB temperature in Figure \ref{maps}, we use this quantity in our
cross-correlation.

\begin{figure}
\centerline{
\psfig{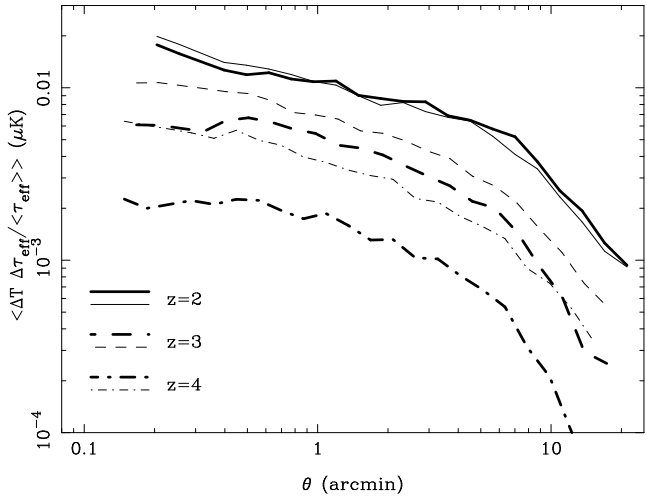}
}
\caption{
Cross-correlation of $|T|$ (for kSZ) or $T$ (for tSZ) and $F/<F>$ as a
function of pixel-pixel distance, for three different redshifts.
Results for tSZ are shown by thick lines and kSZ by thin lines.
\label{corr}
}
\end{figure}

We compute the cross-correlation separately for the tSZ and kSZ
effects. For the kSZ, because the magnitude of the effect can be
positive of negative, we work with $|T_{tSZ}|$. Although this is not
easily done with observational data, in the current section we are
interested in making a 
relative comparison between the magnitude of the tSZ and kSZ
effects, where this is a useful
approach.  In order to do the cross-correlation,
for each \lya\ spectrum we first calculate $\Delta\tau_{\rm
eff}=-(\tau_{\rm eff}-\langle \tau_{\rm eff} \rangle)/\langle
\tau_{\rm eff}\rangle$, where $\langle\tau_{\rm eff}\rangle =
-\ln\langle F \rangle$ and we use the average mean transmitted
flux for all spectra. The cross-correlation we compute is
$\langle\Delta\tau_{\rm eff} \Delta T \rangle$, which has units of
$\mu$K. For the kSZ we use $\Delta T=|T|-\langle|T|\rangle$ and for
the tSZ, $\Delta T=T-\langle T\rangle=T$.  The results for redshifts
$z=2-4$ are shown in Figure \ref{corr}. We have not rebinned the CMB
temperatures to the angular resolution of the spectrum grid, so that
we are able to show results down to angular scales below 1 arcmin.

Unlike the case for the $rms$ $T$ fluctuations, one can see that the
cross-correlation statistic for the kSZ effect depends on redshift,
driven by evolution in the \lya\ forest. On first inspection, this is
somewhat puzzling, as \lya\ \taueff\ fluctuations should be increasing
towards high redshift, as the mean optical depth increases, but
instead they are smaller. This is because we have divided out
$\langle\tau_{\rm eff}\rangle$, and so the fractional fluctuations are
indeed smaller. Measurements based on the flux $F$, instead divide out
$\langle F \rangle$ which is smaller at high redshift (see e.g., Croft
\etal 1998).

This statistic and its spatial dependence could in principle be
measured observationally for the thermal SZ effect.
 We will return to this point in Section 5. 

\subsection{One dimensional distributions}

\label{oned}

\begin{figure*}
\centerline{
\psfig{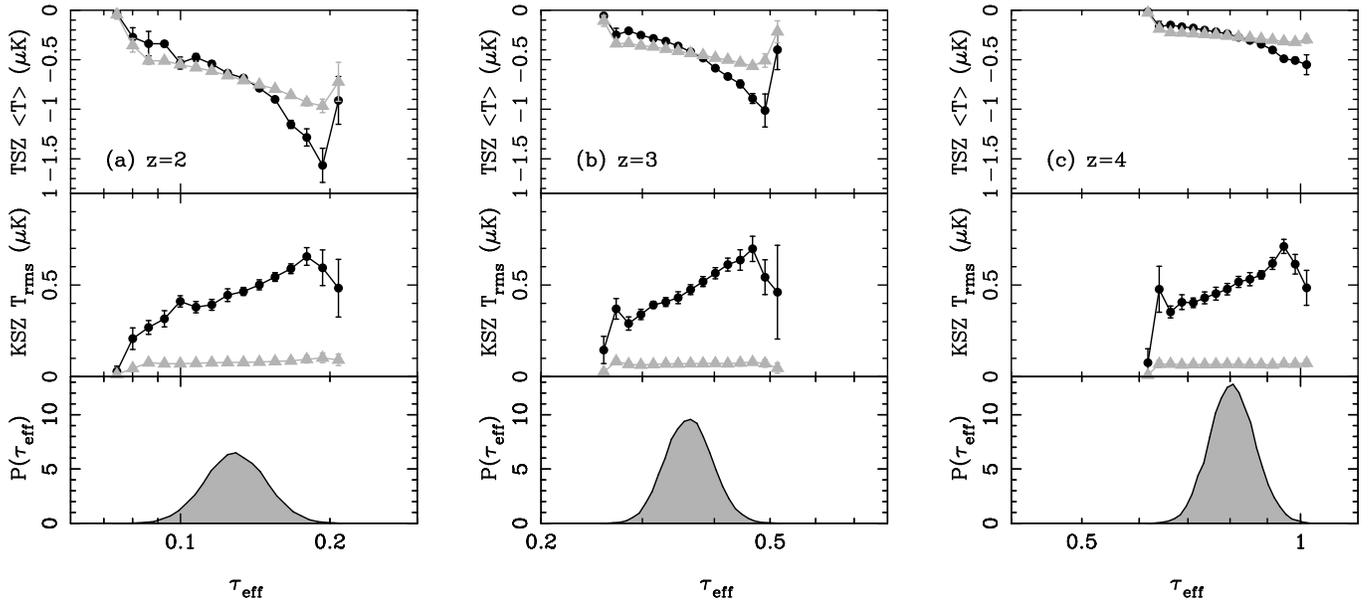}
}
\caption{
Top panels: Dependence of mean temperature from the thermal SZ
effect on $\tau_{\rm eff}=-\log <F>$ for spectra running through the
pixels.  Middle panels: Dependence of rms temperature fluctuation
in pixels (1.5, 1.2 and 1.1 arcmin size for $z$=2,3,4) on $\tau_{\rm
eff}=-\log <F>$ for spectra running through the pixels, for three
different redshifts (left to right). The round symbols which make up
the upper curve in each case show results for pixels, unsmoothed, and
the triangles for maps smoothed with a Gaussian filter of FWHM 13
arcmins, similar to the WMAP W-band resolution. Error bars are from
the scatter between 10 fields in each case (size 95, 77, 69 arcmin for
$z=2, 3, 4$).  Bottom panels: the pdf of $\tau_{\rm eff}=-\log <F>$
for spectra from \lya\ to \lyb.
\label{tauvsrms}
}
\end{figure*}

We have seen from Figure \ref{tvstau} that there is a dependence of SZ
decrement and the \lya\ \taueff\ for spectra in the same pixel. We
will average this relationship over many pixels and bin as a function
of \taueff\ , in order to visualise the mean trend of T vs \taueff\
. This simple statistic is one which we will concentrate on in this
subsection, and use in Section 4 below to compare to observational
data. The mean trend of T vs \taueff\ does depend on the resolution of
the CMB map, and we present results for the same resolution as the
spectrum grid ($64^2$ pixels, each of width 1.2 arcmin at z=3) as well
as smoothing with a Gaussian filter with FWHM 13 arcmin (the WMAP
W-band resolution, see Bennett \etal 2003).  We will also plot the
$rms$ temperature fluctuations in pixels caused by the kinetic SZ
effect.

Figure \ref{tauvsrms} shows our results for three different redshifts,
with the top panels being the tSZ decrement. We can see that the
overall level of the temperature decrement is strongest for $z=2$ and
weakest for $z=4$, as expected. At $z=2$, pixels with \lya\ \taueff\
=0.07 have tSZ temperature decrements of $\sim 0 \mu$K, and the pixels
with \taueff\ =0.2 have decrements $\sim -1.5 \mu$K. Between these two
extremes, the trend of T with \taueff\ is close to linear.  The \lya\
spectra with the very highest levels of absorption, \taueff\
$\simgt0.2$, exhibit a weakening of the decrement, likely owing to
radiative cooling of the gas in these high density regions producing
relatively more neutral hydrogen. The signal smoothed at the WMAP
resolution is somewhat weaker, but still easily measurable from the
simulations.

In the middle panels of Figure \ref{tauvsrms}, we show the kSZ induced
$rms$ temperature fluctuations. They exhibit a rise with increasing
\taueff\ , in accord with the rise in density probed by the
sightlines, again with a turndown for the spectra with the very
highest absorption levels.  As with the $rms$ in circles plotted in
Figure \ref{rmsksz}, the level of fluctuations changes little with
redshift. Smoothing the map to the WMAP resolution does affect the
level strongly, however, reducing the magnitude of the signal by
roughly a factor of ten. As $rms$ fluctuations are already more
difficult to measure in the presence of noise than the mean level,
this means that it is likely that the thermal rather the kinetic SZ
effect will be easiest to target when considering the analysis of
observational data.

One can ask whether the signal from the tSZ and kSZ effects arises in
low or high density gas, and whether we are really sampling the low
density IGM probed by the \lya\ forest. Because our measure of the
\taueff\ - T correlation is angular pixel weighted (we are using a
histogram of pixel values), the fact that most of the sightlines we
use will contain relatively little gas comes into play.  For example,
only a few of the pixels in Figure \ref{maps} pass close to clusters
and groups, and most of the spectra with the lowest 90 percent of
\taueff\ values are likely to pass through only void-like regions in
the \lya\ to \lyb\ interval. This means that while the thermal SZ does
directly probe the mass weighted (or rather electron number-weighted)
IGM temperature, and so is dominated by virialised objects, by
breaking up the signal as a function of \taueff\ we introduce
sensitivity to what is occurring in low density regions.

\begin{figure*}
\centerline{
\psfig{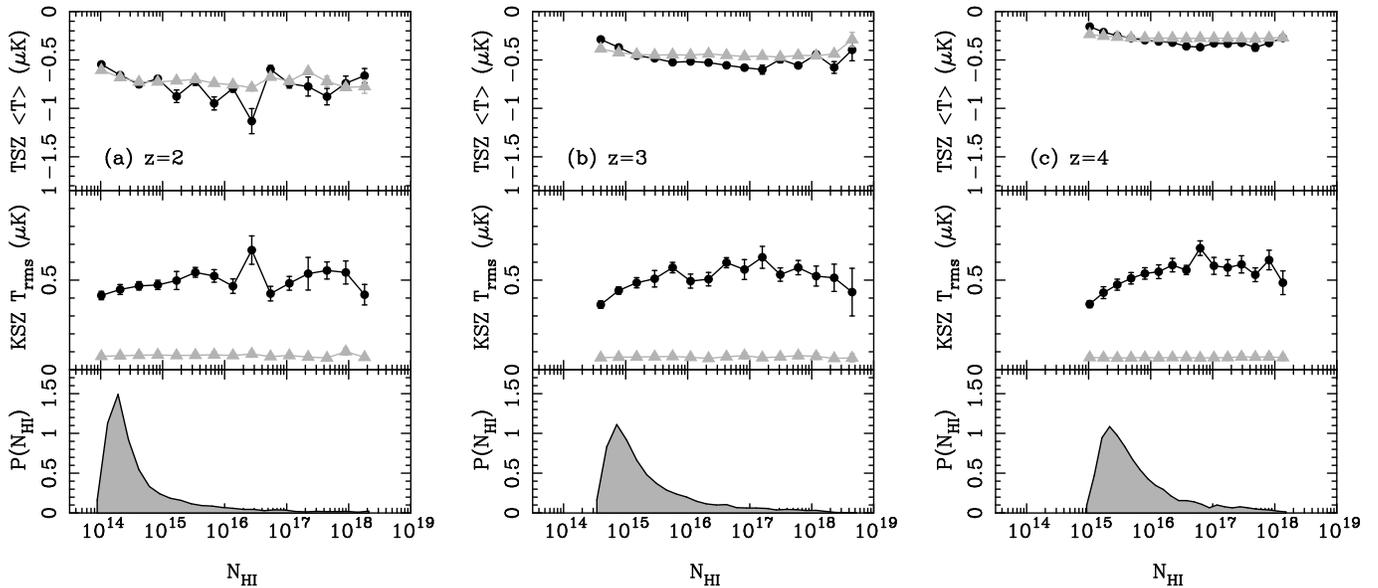}
}
\caption{
Top panels: Dependence of rms temperature fluctuation in pixels (same
size as Figure 7) on NHI for spectra running through the pixels, for
three different redshifts (left to right). The round symbols which
make up the upper curve in each case show results for pixels,
unsmoothed, and the triangles for maps smoothed with a Gaussian filter
of FWHM 13 arcmins, similar to the resolution of the WMAP
satellite. Error bars are from the scatter between 10 fields in each
case (size 95, 77, 69 arcmin for $z=2,3,4$ ).  Bottom panels: the pdf
of $NHI$ for spectra from \lya\ to \lyb.
\label{nh1vsrms}
}
\end{figure*}

\begin{figure}
\centerline{
\psfig{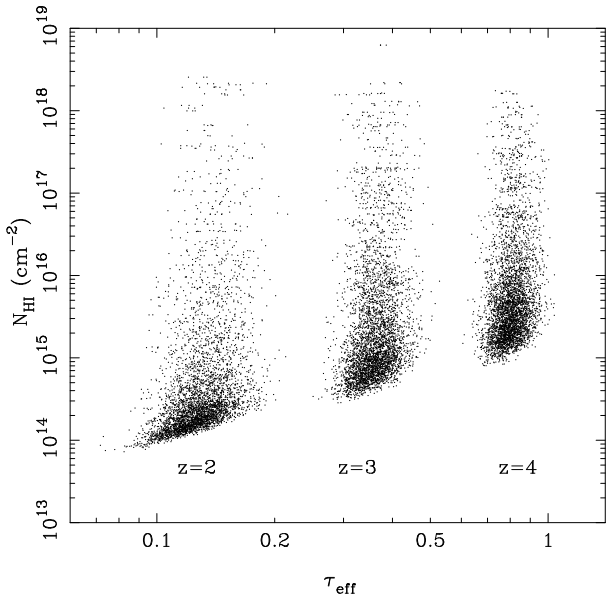}
}
\caption{
$N_{\rm HI}$ vs $\tau_{\rm eff}$ for spectra at three different redshifts.
\label{nh1vsf}
}
\end{figure}

In Figure \ref{maps}, we saw that there appeared to be a much better
correspondence between CMB temperature and \taueff\ than HI column
density.  Nevertheless, from looking at the maps it was not
immediately obvious that the very highest density objects are in fact
associated with large amounts of HI and have a significant SZ
signal. In Figure \ref{nh1vsrms} we plot CMB T vs $N_{HI}$, and see
that there is a slight increase in the SZ signal for sightlines with
$N_{HI}=10^{14}-10^{16}$ cm$^{-2}$, for all redshifts.  The curves,
however, are relatively flat at higher column densities. It appears that
information from saturated regions will not be helpful in carrying out
\lya\ forest CMB correlations, and so there is little need for high
resolution spectra. In order to shed further light on this, in Figure
\ref{nh1vsf} we have plotted values of $N_{HI}$ vs \taueff\ for
individual sightlines, as a scatter plot (as always, we average
over the \lya\ to \lyb\ region in each spectrum). We expect there to
be an envelope, a minimum $N_{HI}$ for sightlines with a given
\taueff\ . This minimum level would correspond to a uniform
distribution of absorbing material (see e.g., Weinberg \etal 1997). We
can see such an envelope, but also a large scatter towards higher
values of $N_{HI}$. The fractional scatter in $N_{HI}$ increases for
larger \taueff\ , so it is not surprising that for high \taueff\
there is little evidence of a trend in Figure \ref{nh1vsrms}.

\section{Limits from SDSS and WMAP data}
\label{wmaps}

\subsection{Method}

The microKelvin fluctuations produced by the \lya\ forest gas are likely
too small to be detected directly in current data (e.g., Bennett \etal
2003, Kuo \etal 2004). However, it is possible that by comparing
observational statistics with simulation results we can come up with
interesting limits on the population of ionised baryons, as well as
estimating the properties of the observations necessary to make a
direct detection. Based on our investigations in the previous
sections, a good strategy will be to measure \taueff\ for the \lya\
forest as well as the CMB temperature in pixels where there are \lya\
spectra.  For the present limited observational study, we choose to
investigate the one-dimensional distributions of these statistics, as
in Section \ref{oned}. We do this for reasons of simplicity, and
because breaking up the signal into sightlines with high and low
absorption levels will enable us to look for an obvious distinct
signature (unlike a full spatial cross-correlation analysis which
would have more discriminating power, but be more difficult to
interpret).

In our analysis, we calculate the mean and rms CMB temperature for
pixels binned using the \taueff\ of the spectra that pass through
them.  We use a jackknife estimator (e.g., Bradley 1982) to compute
the error bars, breaking up the data into 50 subsamples.  We have also
computed error bars by randomly rotating the CMB data with respect to
the \lya\ forest data and find them to be broadly consistent.

\subsection{\lya\ forest data}

We use the \lya\ forest of quasar spectra taken from Data Release 3 of
the Sloan Digital Sky Survey (Abazajian \etal 2005). Our data sample
consists of 3488 quasar spectra with quasar redshifts greater than
$2.4$, pruned to remove Broad Absorption Line quasars. The
particular dataset was compiled from the publically available data by
Scott Burles and kindly made available to us (Burles, private
communication).  Owing to the relatively high low-redshift limit for
the quasars, the spectra consist of the full \lya\ to \lyb\ region of
the forest, lying above the atmospheric cutoff at $\sim 3200$ \AA\ .
This will be useful in order to be consistent with what was done in
the simulations.

The one-dimensional distribution of \taueff\ requires that the data be
divided into broad bins (we will use five bins). In order to calculate
\taueff\ for each spectrum, we need to fit the continuum over the full
\lya\ to \lyb\ region.  Historically, the averaged value of the mean
transmitted flux from many spectra has been used to calculate a global
value of \taueff\ (e.g., Kirkman \etal 2005).  Fluctuations in
\taueff\ for entire spectra have not been estimated, however,
presumably owing to problems with calculating a mean value on large
scales where continuum fluctuations are seen to be comparable to
fluctuations in the absorption (see for example the power spectrum of
the raw spectrum in Hui \etal [2001] or McDonald \etal [2004]). The
largest scale on which fluctuations intrinsic to the \lya\ forest have
been measured is 153 Mpc at redshift $z=1.9$ by Tytler \etal (2004),
who find that Lyman limit systems and metal lines make large
contributions to fluctuations in the mean flux at this relatively low
redshift.  In our case, we will be working at $z=3$, where these
effects should be less important. Nevertheless, we expect that there
will be a large contribution to the variation in \taueff\ from
sightline to sightline owing to continuum estimate errors as well as
noise in the spectra and non-\lya\ absorption. We leave the question
of how best to fit the continuum on the largest scales to other work
(e.g., Suzuki \etal 2005), and in the meantime, we will add a Gaussian
distribution of errors to our simulation measurement to account for
these effects when comparing our simulations with observations. The
relatively large bin size should serve to mitigate effects of \taueff\
errors in this first illustrative comparison.

We fit the continuum using the often employed polynomial fitting
method (e.g., Croft \etal 1998).  A low order polynomial (3rd order in
our case) is fit to the \lya\ forest region. We avoid the \lya\ and
\lyb\ emission lines by restricting our fit to the rest wavelength
range 1030-1190\AA. We then remove all points more than 2 $\sigma$
below the continuum and refit, iterating this procedure until
convergence is achieved. Averaging over all \lya\ pixels in our
dataset, we find a mean redshift $z=2.91$. At this stage we divide out
the continuum and compute a value of $F$. For all spectra, we find
$F=0.8$.  Measurements of the mean flux and global value of \taueff\
from higher resolution datasets better suited to this measurement find
a lower value of $F$ (e.g., Schaye \etal 2004, Kirkman \etal 2005).
Our measurement of $F$ is on average biased high because we miss
absorption in flatter low density regions around the continuum owing
to our low resolution. We correct for this bias in a crude way by
multiplying the value of the fit continuum in each case by
$1/0.85$. This simple correction is adequate for our current
purposes, having been set so that we reproduce the global \taueff\
value of Schaye \etal (2004) at $z=3$. We make the assumption that any
random variations in this bias from spectrum to spectrum will simply
act as an additional source of noise which we can model in our
simulation data.

\begin{figure}
\centerline{
\psfig{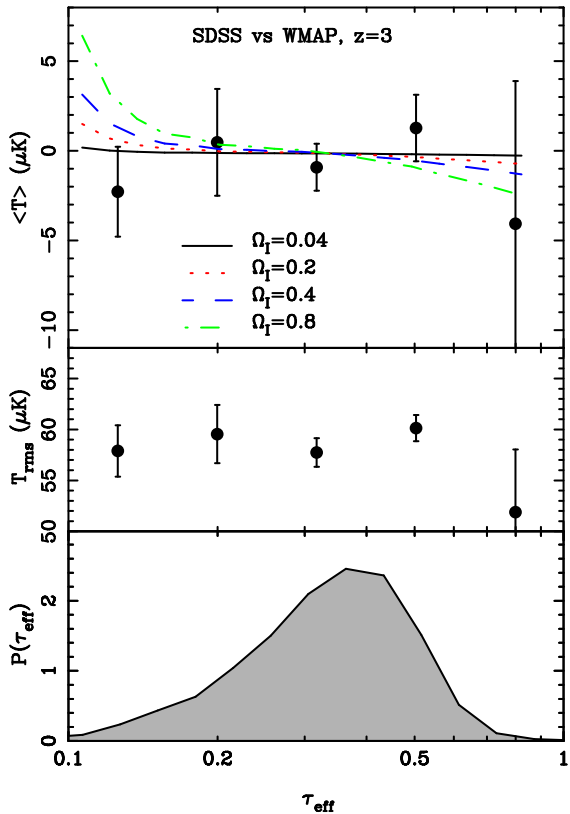}
}
\caption{
SDSS $\tau_{\rm eff}$ vs WMAP $T$ for $z=3$. We show jackknife errors
(10 subsamples). We have subtracted off a ring for background from
15-25 arcmin around the pixel that the spectrum is in. The lines show
predictions from LCDM after broadening the $\tau_{\rm eff}$ values with
observational errors. The solid line is for the LCDM baryon density,
dotted for LCDM $\times$ 5 (i.e. $\Omega_{b,I}=0.4$) and dash-dotted
LCDM $\times$ 20 (i.e. $\Omega_{b,I}=0.8$).
\label{wmap}
}
\end{figure}

In the bottom panel of Figure \ref{wmap} we have plotted a histogram
of the \taueff\ values derived from the SDSS spectra. The mode of the
distribution is \taueff\ =0.35, the same as the simulation results at
$z=3$ plotted in Figure \ref{tauvsrms}. In the observational data,
there is a tail extending down to $\tau_{\rm eff} \sim 0.1$ and up to
$\tau_{\rm eff} \sim 1$ which owes to noise and the continuum fitting
uncertainties described above.

\subsection{CMB data}

We make use of the WMAP first year data (Bennett \etal 2003) in our
analysis. For the illustrative application of \lya\ -CMB
cross-correlation in this paper, we employ the foreground cleaned
version of the data presented by Tegmark \etal (2004). The CMB
temperature is given in HEALPix format ($12\times512^{2}$) pixels, at
the resolution of the WMAP W-band data (13 arcsec FWHM). 

The Tegmark \etal map is chosen by us because it combines the 
5 WMAP channels into a map with the highest angular
resolution possible.
However, because it is derived from channels with different
frequencies and the thermal SZ effect has a weak 
frequency dependence, the map will have a scaled value of the SZ effect present,
dependent on the weights used to combine the channels. In order to understand
what impact this will have, we can use Equation 3 to examine the
thermal SZ  temperature fluctuations in the different bands. Equation
3 can be written in the form $\Delta T/T=f(\nu) \times y$, and at the 
lowest frequencies, $f(\nu) \rightarrow -2.0$.
The five WMAP
channels are centered at frequencies of 23, 33, 41, 61 and 94 GHz. The
values of $f(\nu)$ for the channels are then $-1.97, -1.94, -1.91, -1.81$ 
and $-1.56$, respectively. The weights for the Tegmark map vary spatially
and are not publically available, but one can see the maximum difference of 
the effect compared to our Rayleigh Jeans approximation ($f(\nu)=-2.0$) by
considering the 94 GHz W-band map on its own, which gives 
approximately a $22 \%$ 
reduction in the predicted tSZ signal. If we instead assume that the weights
will have a similar effect to those used by the WMAP team itself
for their linearly combined map, we find by
summing over the 5 bands that $f_{\rm effective}=\sum_i^{5} W_{i}
f_{i}(\nu) = -1.79$, or a $\sim 10\%$ reduction, The effect of 
frequency dependence is therefore small enough and uncertain enough
that we neglect it in our illustrative application to observational data.
In the future, however, as we have mentioned before, the frequency dependence
can be used to better extract the signal. We will return to this point. 

 For each
pixel centered on one of the SDSS quasar spectra, we extract the
corresponding CMB temperature of that pixel. In order to crudely
mitigate the effects of primary CMB anisotropies, which act as
noise for our signal, we subtract a local background consisting of the
CMB temperature averaged over all pixels in an annulus of
inner radius 15 arcmin and outer radius 25 arcmin, centered on the
pixel in question.  Our motivation for this background subtraction is
the fact that the SZ signals fall off rapidly on large scales
(e.g. see Figure \ref{corr}), unlike the CMB primary signal. Because
we will only be computing an upper limit to the SZ signal, this
subtraction is a conservative assumption.

The presence of CMB primordial anisotropies will produce fluctuations
on large scales. For example, there will be coherent fluctuations on
the 100 arcmin scales which our simulated sky maps cover. The annulus
subtraction technique represents a crude way to deal with this. As we
shall see below, the noise level on the pixel scale is sufficiently
large with the WMAP data that we are not limited by the primary
fluctuations in our analysis. However, in the future, with better data
it will become imperative to model and remove the effect of the
primordial anisotropies.  This could be done with matched filtering
(e.g., Schaefer \etal 2004), and the technique tested with simulations
of the CMB sky which include large-scale anisotropy, foregrounds,
noise and an instrumental beam response. We leave this for future
work, and in Section \ref{disc} we will discuss other more
sophisticated ways to deal with the CMB data.

\subsection{Results}

We find the $rms$ temperature fluctuations in all 3488 pixels
containing spectra to be $60 \mu$K.  In the middle panel of Figure
\ref{wmap} we show the $rms$ temperature in each of 5 \taueff\
bins. The error bars range from 1.3 to $6.1 \mu$ K. This is close to
the Poisson error estimate in each case, which is not surprising
because the pixels are sampled quite sparsely from the data.  Given
that we expect $\sim 1 \mu$K $rms$ fluctuations at most from the
kinetic SZ effect, it is clear that this particular dataset will fall
short of giving us any meaningful constraint from the kSZ.  Because
the kSZ signal will add in quadrature to the $rms$ from the noise and
CMB primary signal, the highest \taueff\ bin might have a total $rms$
of $\sqrt{60^{2}+1^{2}}\mu$K and the lowest $\sqrt{60^{2}-1^{2}}\mu$K,
a difference of 0.02 $\mu$K. In practice, because the CMB data is
smoothed, we expect at least a factor of 10 less signal (see the
middle panel Figure \ref{tauvsrms}). The kSZ signal is therefore at
least 500 times too small to be detectable in this way with these
data.

If instead we turn to the thermal SZ signal, the situation is much
more promising.  As a mean temperature rather than a variance, it is
easier to constrain this signal from noisy data. In the top panel of
Figure \ref{wmap} we show the mean temperature in pixels for the 5
bins of \lya\ \taueff\ .  Owing to the background subtraction, the
mean value of the pixels is zero. When comparing to the simulation
data, we will renormalise so that the same is true. The observational
error bars vary from 1.3 to 8.0 $\mu$K per bin.

From Figure \ref{tauvsrms}, we expect the mean temperature decrement
from the thermal SZ effect at redshift $z=3$ to exhibit a total
variation of $\sim 0.6 \mu$K amongst the pixels with highest and
lowest \lya\ forest \taueff\ . This is for CMB maps at the WMAP
resolution, but unfortunately not including noise in the \lya\
\taueff\ estimates. As detailed in Section 4.2 above, we model the
effects of noise in the spectra and continuum fitting uncertainties by
adding a Gaussian-distributed error to the \taueff\ from simulations.
We add a \taueff\ value randomly drawn from a normal distribution with
standard deviation $\sigma=0.1$.  Doing this to the simulated values
of \taueff\ enables us to reproduce the width of the histogram of
observed \taueff\ values.  In the top panel of Figure \ref{wmap} we
plot the curve of simulation CMB temperature $\langle T \rangle$
vs. simulated \taueff\ after this noise has been added. The line which
lies closest to the $\langle T \rangle=0$ axis is this prediction. As
we can see, the predicted variation of $\langle T \rangle$ with
\taueff\ is barely noticeable on this scale, indicating that tight
constraints lie only in the future with larger observational datasets.

Nevertheless, the comparison is interesting because it constitutes an
attempt to detect the ionised baryons at $z=3$, and the sensitivity
increase required for success if our current cosmological models are
correct is only about one order of magnitude. In order to place an
upper limit on the temperature and density of ionised baryons using
existing data, one would ideally use a grid of simulations with
different values of the baryon density, with the amplitude of matter
fluctuations normalised so that their clustering and other properties
are consistent with observations (e.g. Spergel \etal 2003,
Tegmark \etal 2004).  In this manner, both the clustering of baryons
and the variation of mean IGM temperature with $\Omega_{b}$ could be
marginalised over to yield a constraint on $\Omega_{b}$ from ionised
baryons, $\Omega_{\rm b,I}$. The effects of varying $\Omega_{b}$ on
structure at high redshift were explored with a suite of simulations
by Gardner \etal (2003).  These authors found for example that \lya\
forest absorption is virtually unaffected by changing $\Omega_{b}$ (by
a factor $\sim6$), as long as the ionising background intensity is
adjusted to reproduce the same value of \taueff\ .

In the present case, because of the large error bars, the value of
$\Omega_{\rm b,I}$ consistent with observations is likely to be so
high that a simpler and more crude approach than a full simulation
grid is all that is warranted. In order to model the effect of
increasing $\Omega_{b}$ we simply multiply the simulation value of SZ
$\langle T \rangle$ by a constant factor, assumed to be equivalent to
multiplying $\Omega_{b}$ by the same factor. We do not adjust the
\lya\ observables, imagining that any increase in baryon density can
be cancelled out by the presence of a UV background radiation field
higher in intensity by the same factor.  Moreover, we assume that
increasing $\Omega_{b}$ does not affect the amplitude of matter
fluctuations and that the temperature of the IGM stays the same as in
the $\Omega_b=0.04$ simulation. A model with a very large value of
$\Omega_{b}$ is likely to have many other differences related to gas
physics and cooling, such as the number of collapsed baryonic
objects. Additionally, of course, it will be incompatible with Big
Bang Nucleosynthesis constraints (e.g., Burles and Tytler 1998), and
measurements of $\Omega_{b}$ from the CMB acoustic peaks (Spergel
\etal 2003).  None of these problems concern us here, as our
comparison to simulations is essentially illustrative.

Under these assumptions, a $\chi^{2}$ fit of the simulation data to
the WMAP/SDSS data, both binned in the same manner, yields
$\Omega_{\rm b,I} < 0.55$ at $90\%$ confidence and $\Omega_{\rm b,I} <
0.80$ at $95\%$ confidence. The limits on the fraction of the critical
density in ionised baryons at $z=3$ are therefore very weak.  This is
as one would expect from looking at Figure \ref{wmap} where we show
the simulation curves for various values of $\Omega_{\rm b,I}$.  It is
possible to conclude that the Universe is not closed by ionised
baryons, but not much else. This situation is likely to be much
improved with future observational data, as we will discuss below.

Much of the statistical power in the measurement comes from the bin
with the lowest value of \taueff\ .  Most of the pixels have moderate
values of \taueff\ but these pixels in the lowest bin also have low SZ
decrement amplitudes. In this manner, the CMB-\lya\ forest
cross-correlation can be said to be probing the low density IGM.

\section{Summary and Discussion}

\subsection{Summary}

Using a cosmological hydrodynamic simulation we have investigated the
CMB temperature fluctuations caused by the diffuse IGM at redshifts
$z=2-6$, also responsible for the \lya\ forest.  Our main findings can
be summarised as follows:\\
\\
\noindent(1) At redshift $z=2$, the kinetic Sunyaev-Zel'dovich effect
from gas within the \lya\ to \lyb\ region of the forest is predicted
to cause CMB fluctuations of amplitude 0.8 $\mu$K rms averaged in a
top-hat filter of angular radius 1 arcmin. This remains roughly constant
for higher redshifts.\\
\\
\noindent (2) The thermal SZ at redshift $z=2$ yields an rms T of 2.4
$\mu$K. At higher redshift, the relative fluctuation owing to the
thermal SZ compared to kinetic SZ declines as expected because of the
decreasing mean temperature. However, there is still a substantial
$T_{\rm rms}$ at higher redshift, 0.4 $\mu$K at $z=4$.\\
\\
\noindent(3) There is a correlation between \lya\ forest properties of
quasar spectra with CMB temperature fluctuations caused by both the
kinetic and thermal SZ effects. In particular, there is a strong
relation between the mean effective optical depth \taueff\ averaged
over the entire \lya\ forest region and the CMB temperature.  Pixels
(of size $\sim$ 1 arcmin) in the top 10 percentile of \taueff\ are
predicted to have a mean CMB temperature 1.5 $\mu$K lower than the
bottom 10 percentile at $z=2$, owing to the thermal SZ effect.  The
quantity \taueff\ is relatively sensitive to density fluctuations in
the diffuse IGM.\\
\\
\noindent(4) There is, however, little correlation between the total
HI column density in spectra and the CMB temperature, with no relation
at all seen for column densities $N_{HI}>10^{16} cm ^{-2}$. This
statistic is most sensitive to the neutral hydrogen density in
saturated regions, indicating that SZ effects correlate best with
relatively diffuse unsaturated gas.\\
\\
\noindent(5) We have compared the correlation of \lya\ \taueff\ with
CMB T seen in simulations of the $\Lambda$CDM cosmology with
observational data from the Sloan Digital Sky Survey and WMAP. Because
the kinetic SZ signal and pixel noise add in quadrature, no
constraints are possible on the IGM velocity field with the present
data. However, the sensitivity is within an order of magnitude of that
necessary to detect the predicted thermal SZ effect from the \lya\
forest. We are able to place a weak limit on the density in ionised
baryons at $z=3$ of $\Omega_{B,i} < 0.80$ at $95 \%$ confidence. Below
we outline future developments which would make a direct detection
possible.

\subsection{Discussion}

\label{disc}

In the present paper, we have examined the relationship between \lya\
forest absorption and the SZ effect in general, without the
expectation that a detection of a correlation can be made from current
data. In the future, the quality and size of datasets will increase,
and it will become worthwhile to plan a strategy for the
cross-correlation that makes the best use of the data. Two obvious
improvements could be made in the future. In our illustrative
comparison of WMAP and SDSS data we did not use the full angular or
frequency information, both of which would help with the sensitivity
of the test.

As we have seen that the thermal SZ effect offers the best hope for
detection of the ionised baryons, making use of the spectral
difference between the SZ decrement and the primary CMB signal will
enable easier detection of the effect.  One could imagine fitting the
\taueff\ -T correlation in different CMB frequency bands to do
this. For example, in the work of Huffenberger \etal (2004), a method
was derived for combining the WMAP channels in an optimal manner to
remove the primary anisotropy, effects of noise and contaminating
point sources, enabling the thermal SZ signature to be constrained
directly. In this paper we have concerned ourselves with predictions
for the Rayleigh Jeans part of the spectrum, which is reasonable given
the large observational uncertainties in present data. We note that at
higher frequencies the thermal SZ signal becomes positive (for example
at 330 GHz the magnitude is the same as the Rayleigh Jeans signal but
with the opposite sign). Bolometers tend to be more sensitive than
radiometers at present, and they operate at frequencies $>100$ GHz.
These differences could be used profitably to increase the signal to
noise of any detection.  Also, with the WMAP data, we have limited our
analysis to single pixels.  Making use of the full spatial
cross-correlation in our detection of the signal will be useful when
CMB datasets with higher angular resolution become available.
Another possibility could be to use wavelets to isolate the
relevant physical scales in a future analysis (see e.g., Vielva \etal
2005).

One can ask whether emission from the QSOs themselves might create a
signal, as the SED of QSOs in the microwave band is not well
constrained (Perna \& Di Matteo 2000, White \& Majumdar 2004).  Apart
from the possible intrinsic microwave emission of QSOs, there will
also be an intrinsic thermal SZ effect from galaxy clusters which host
QSOs.  Neither of these will necessarily bias any measurement of the
\lya\ -CMB correlation unless QSOs with more \lya\ absorption in their
spectra have more intrinsic microwave emission.  This seems unlikely
in the former case, although in the case of an SZ signal associated
with clusters in which the QSOs lie one could argue that this is part
of the signal that we are trying to measure.  Otherwise, it could be
removed if we so desire by not using the part of the spectrum closest
to the QSO.

The method we have outlined in this paper is one way to potentially
detect ionised baryons in the IGM at high redshift. In general, one
would like to know if there are any other methods that can be used. At
lower redshifts, for example ($z<1$), the ionised baryons are mainly
collisionally ionised, with such a low neutral fraction that they
cannot be seen in the \lya\ forest.  Their higher temperature, however,
means that this Warm Hot Intergalactic Medium (WHIM, see e.g., Cen
\etal 1995, 1999, Dav\'e et al 2001) can potentially be detected
by looking for X-ray absorption (e.g., Fang \etal 2002) or emission
(Yoshikawa \etal 2004, Fang \etal 2005) or perhaps in \lya\ emission
(e.g. Furlanetto et al. 2003, 2005).  The thermal SZ signal at low
redshifts is also one way to detect the WHIM. At higher redshifts and
lower IGM temperatures considered in this paper however, the X-ray
signal (see e.g., Croft \etal 2001) is not likely to be observable.

The overall $y$-distortion in the CMB spectrum from the hot IGM should
eventually be detectable. Limits from the COBE FIRAS instrument (e.g.,
Mather \etal 1990) are $y < 2.5 \times 10^{-5}$ at $95 \%$ confidence,
more than an order of magnitude above the mean $y$ caused by the
high-$z$ IGM.  However, separating out the contribution from the high
$z$ IGM will be difficult without a spatial correlation of the type we
have proposed here. One could also use other tracers of the high $z$
density field for a cross-correlation, such as Lyman-break galaxies,
or QSOs themselves, although their connection to the IGM would be less
easy to interpret.

Another interesting question to ask is how likely it is that we will
be able to constrain the high $z$ IGM velocity field, given that the
kinetic SZ signal from the photoionised IGM is expected to be
relatively important (Loeb 1996). Because of the possible +ve or -ve
nature of the effect, depending on the velocity, we cannot use a mean
signal, unlike the thermal SZ. By using an $rms$ fluctuation as our
statistic, we therefore need a noise level which is extremely
low. Since the spectrum of the kSZ is the same as the CMB, the
primary fluctuations become even more important sources of noise than
with the thermal effect.  Given that the rms fluctuations on scales of
1 arcmin from the kSZ are predicted to be $\sim 5\mu$K at $z=2$, one
would need the ``noise'' from primary anisotropies and other sources
to be less than this in order to detect a signal.

The thermal SZ is more promising. Future experiments which will be
useful for improving constraints on $\Omega_{b,I}$ include the Planck
satellite mission\footnote{http://planck.esa.int},
 the Sunyaev Zel'dovich Array (SZA\footnote{http://astro.uchicago.edu/sza}),
 the Atacama Cosmology Telescope (ACT, see Kosowsky 2003), and the South
Pole Telescope (SPT\footnote{http://astro.uchicago.edu/spt}).
  Given the expected noise characteristics and
angular resolution/sky coverage of these instruments, we expect
substantial improvements over the constraints derived using WMAP in
this paper. We have seen in the WMAP case (Section \ref{wmaps})  that we are
limited both by the noise level ($\sim60 \mu $K rms) and the
resolution of the data. As an example, the ACT angular resolution is
projected to be $\sim 1$ arcmin with an $rms$ detector noise level of
$\sim 2 \mu$K per pixel in the 145 GHz band. Even without allowing for
the higher resolution or using the full angular cross-correlation
information, one would therefore expect to extract the same level
\lya\ forest thermal SZ signal using $(60/2)^{2}$ fewer spectra
(assuming the primary anisotropy can be filtered out in frequency
space). The ACT signal 
will be lower by about $50\%$ at 145 GHz  than the Rayleigh Jeans
prediction, but the other frequency bands can be used to increase
the significance of the detection.
 In order to yield a 2-$\sigma$ detection of $\Omega_{b}$,
using pixel-pdf statistics, as in this paper, we estimate that $\sim$
300 quasar \lya\ forest spectra would be needed in the ACT survey area,
assuming the BBN value of $\Omega_{b} h^{2}$ (0.019; Burles \& Tytler
1998) Using angular cross-correlation information would increase the
significance of the detection.  The lack of overlap between the ACT
fields and the SDSS would necessitate a southern quasar survey,
although it should be borne in mind that only extremely low resolution
spectra are needed.

We note that in our analysis we have assumed that the ratio of neutral
to ionised hydrogen is that expected if the ionising background
radiation field does not vary spatially in  intensity. If there
are coherent spatial fluctuations in the photoionisation rate ($\Gamma$ in 
Equation 1) then the correlation between \lya\ optical
depth and CMB temperature decrement will be affected by an additional
source of scatter. This is likely to be small, however
with the spatial cross-correlation reduced by $5\%$ or less (see
e.g., Croft 2004) and could in principle be calibrated out using theoretical
modelling in the manner of McDonald \etal 2005.

Finally, one can ask why it is important to try to detect the ionised
component of the baryons at high redshift when the neutral hydrogen
tracer has already been well measured (e.g., Rauch \etal 1997).  One
reason is that a detection would act as a consistency check and enable
us to verify directly the baryon inventory at these redshifts. Also,
at the moment, the best constraints on the baryon density in observed
structures (if we set aside for the moment relatively indirect CMB and
Big Bang Nucleosynthesis, BBN measurements e.g., Spergel \etal 2003,
Walker \etal 1991) comes from looking at the mean absorption level in
the \lya\ forest (e.g., Rauch \etal 1997, Weinberg \etal 1997, Hui
\etal 2002, Tytler \etal 2004).  However, these measurements are
degenerate with constraints on the ionising background radiation
intensity $J$.  Often, $\Omega_{b}$ from BBN or CMB is assumed and
then the \lya\ forest measurement is used to constrain the ionising
radiation intensity (e.g., Kirkman \etal 2005).  If we use actual
measurements of $J$ (from e.g., the proximity effect e.g., Scott \etal
2000) then constraints on $\Omega_{b}$ directly seen in the \lya\
forest become weaker ($J$ is only known to within roughly a factor of
two). Measurements of the ionised baryon fraction such as those we
propose may therefore become competitive.

\section*{Acknowledgements}
We thank Scott Burles for providing the modified SDSS DR3 dataset used
in this paper and also Volker Springel for allowing us to use the 
cosmological simulation data. RACC thanks Tiziana Di Matteo, Volker Springel,
Jeff Peterson and Xuelei Chen for useful
discussions and the hospitality of Simon White and the Max Planck
Institute for Astrophysics, where this work was largely carried out.
RACC acknowledges partial support from NASA ATP contract ATP05-39.

\end{document}